\documentclass[11pt,onecolumn]{article}

\usepackage[a4paper,margin=2.4cm]{geometry}  
\usepackage[T1]{fontenc}
\usepackage[utf8]{inputenc}               \usepackage{lmodern}                      
\usepackage{float}
\usepackage{placeins}
\usepackage{multirow}
\usepackage{multicol}
\usepackage{array}
\usepackage{makecell}
\usepackage{geometry}
\usepackage{graphicx}
\usepackage{physics}
\usepackage{textcomp}
\usepackage{amssymb}
\usepackage{siunitx}
\usepackage{etoolbox} 
\usepackage{authblk}
\usepackage[square,numbers,sort&compress]{natbib}
\usepackage[dvipsnames]{xcolor}
\usepackage[
  colorlinks=true,
  linkcolor=MidnightBlue, 
  citecolor=MidnightBlue, 
  urlcolor=MidnightBlue,
  pdfencoding=auto
]{hyperref}
\usepackage{orcidlink}

\def\hlwords#1 #2\relax{%
  \colorbox{yellow}{\strut#1}%
  \ifx\relax#2\relax
  \else\space\hlwords#2\relax
  \fi
}
\AtBeginEnvironment{thebibliography}{%
  \small
  \setlength{\bibsep}{1pt}
}

\DeclareMathAlphabet{\mathcal}{OMS}{cmsy}{m}{n}

\title{Cryogenic RF-to-Microwave Transducer based on a DC-Biased 
Electromechanical System}

\newcommand{\equalcontrib}{\protect\footnotemark[\value{footnote}]}

\author[1]{Himanshu Patange\,\orcidlink{0009-0008-7624-1296}}
\author[1]{Kyrylo Gerashchenko\,\orcidlink{0009-0000-8874-1249}}
\author[1,2]{Rémi Rousseau\,\orcidlink{0009-0009-5630-3047}}
\author[1]{Paul Manset\,\orcidlink{0000-0001-7825-6354}}
\author[1]{Léo Balembois\,\orcidlink{0009-0008-0666-7312}}
\author[3,4]{Thibault Capelle\,\orcidlink{0000-0003-2732-8332}}
\author[1]{Samuel Deléglise\,\orcidlink{0000-0002-8680-5170}\thanks{These authors jointly supervised the work}}
\author[1, 5]{Thibaut Jacqmin\,\orcidlink{0000-0002-0693-4838}\equalcontrib\thanks{Corresponding author: \texttt{thibaut.jacqmin@lkb.upmc.fr}}}

\affil[1]{Laboratoire Kastler Brossel, Sorbonne Université, CNRS, ENS–Université PSL, Collège de France, 4 place Jussieu, 75005 Paris, France}
\affil[2]{Alice \& Bob, 53 Bd du Général Martial Valin, 75015 Paris, France}
\affil[3]{Niels Bohr Institute, University of Copenhagen, Blegdamsvej 17, 2100 Copenhagen, Denmark}
\affil[4]{Center for Hybrid Quantum Networks (Hy-Q), Niels Bohr Institute, University of Copenhagen, Copenhagen, Denmark}
\affil[5]{Institut Universitaire de France (IUF), 1 rue Descartes, 75231 Paris, France}

\begin{document}

\maketitle

\begin{abstract}
We report a two-stage, heterodyne rf-to-microwave transducer that combines a tunable electrostatic pre-amplifier with a superconducting electromechanical cavity.  A metalized Si$_3$N$_4$ membrane (3~MHz frequency) forms the movable plate of a vacuum-gap capacitor in a microwave LC resonator.  A dc bias across the gap converts any small rf signal into a resonant electrostatic force proportional to the bias, providing a voltage-controlled gain that multiplies the cavity’s intrinsic electromechanical gain.  In a flip-chip device with a 1.5~µm gap operated at 10~mK we observe dc-tunable anti-spring shifts, and rf-to-microwave transduction at 49~V bias, achieving a charge sensitivity of 87~\textmu e/$\sqrt{\mathrm{Hz}}$ (0.9~nV/$\sqrt{\mathrm{Hz}}$).  Extrapolation to sub-micron gaps and state-of-the-art $Q>10^8$ membrane resonators predicts sub-200 fV/$\sqrt{\mathrm{Hz}}$ sensitivity, establishing dc-biased electromechanics as a practical route towards quantum-grade rf electrometers and low-noise modular heterodyne links for superconducting microwave circuits and charge or voltage sensing.

\end{abstract}

\section{Introduction}\label{sec:intro}

Bridging the gap between radio-frequency (rf) and microwave-frequency electrical signals is a longstanding challenge in quantum electronics and precision sensing.  If an rf signal can be up-converted without adding noise, it can be processed by near-quantum-limited amplifiers that are available in the microwave frequency band.  In a typical electro-mechanical transducer, an rf charge modulation drives a mechanical mode into motion. The motion is then converted into a microwave sideband that can be measured with sub-nV/$\sqrt{\mathrm{Hz}}$ sensitivity. Such a device can also be used as a sensitive electrometer.

The seminal electrometer of Knobel and Cleland exploits a radio-frequency single-electron transistor to measure the motion of a suspended beam and achieves a charge sensitivity of 12~\textmu e/$\sqrt{\mathrm{Hz}}$~\cite{Knobel2003}.  Subsequent transport-based devices—rf-SETs~\cite{Schoelkopf1998, Blencowe2000, Brenning2006, Aassime2001, Angus2008, Korotkov1999}, rf-QPC~\cite{Cassidy2007}, quantum dots~\cite{Viennot2014, Lu2003, Volk2019}, gate-based devices~\cite{Gonzalez2015}, and fluxonium qubits~\cite{Najera2024, Gerashchenko2025}—have pushed sensitivities below 10~\textmu e$/\sqrt{\mathrm{Hz}}$. A parallel track replaces the electronic detector with an optical or microwave cavity coupled to a high-Q mechanically compliant capacitor. In such schemes, one exploits the high susceptibility of the resonant mechanical system to probe an external force, such as an electrostatic modulation, with high sensitivity. Cavity-enhanced microwave read-out of a nanobeam was obtained at room temperature~\cite{Faust2012}. Additionally, ground-state cooling and imprecision at the standard quantum limit have been achieved in both microwave~\cite{Teufel2011, Rocheleau2010, Seis2022} and optical~\cite{Chan2011, Rossi2018, Peterson2016, Qiu2020} domains, while ultracoherent Si$_3$N$_4$ membranes with Q>10$^9$~\cite{Cupertino2024, Bereyhi2022, Ghamidi2018, Seis2022} for mechanical modes in the 100~kHz range, now suppress thermomechanical force noise to the zeptonewton level~\cite{Seis2022, Bereyhi2022}.  Meanwhile, cavity electro-optomechanical converters have linked microwave and optical photons~ \cite{Andrews2014,Higginbotham2018,Bagci2014,Forsch2020,Mirhosseini2020}.

In most cavity-based converters the microwave cavity alone supplies the gain. The linearized coupling grows with the intracavity photon number, so that achieving higher conversion efficiency requires strong pumps—at the cost of photon shot-noise or heating. Here we report on a two-stage architecture that sidesteps that trade-off by pre-amplifying the rf signal electrostatically before it reaches the cavity. More precisely, we propose a heterodyne rf-to-microwave converter based on a dc-biased membrane. It consists of a metalized Si$_3$N$_4$ membrane whose fundamental drum mode lies near 3 MHz.  The membrane forms one plate of a vacuum-gap capacitor in a superconducting lumped-element LC resonator at 6 GHz, such that its motion modulates the microwave resonance frequency.  A dc bias applied across the gap converts any small rf signal on the same line into a time-varying electrostatic force that scales linearly with the bias voltage~\cite{Bagci2014, Unterreithmeier2009, Viennot2018, Gerashchenko2025}. This electrostatic stage adds a voltage‑tunable pre‑amplification to the RF signal before it is up-converted by the LC microwave resonator. Because the two stages act in series, their gains multiply, boosting the overall RF‑to‑microwave conversion. A related scheme was demonstrated at room temperature in Ref.~\cite{Bagci2014}, where a sub‑MHz rf voltage is converted to an optical carrier with a voltage sensitivity of a few pV/$\sqrt{\mathrm{Hz}}$, 
and further experimental and theoretical developments along similar lines have been reported in
Refs.~\cite{MoaddelHaghighi2018,Bonaldi2023}.
  In the present cryogenic implementation, we demonstrate a flip-chip assembly that realises a 1.5 µm vacuum gap, and observe dc-controlled frequency shifts on the mechanical modes — a static effect known as anti-spring effect~\cite{Unterreithmeier2009, Bagci2014, Gerashchenko2025}.  With a 49~V bias, the device  achieves a charge sensitivity of 87~\textmu e/$\sqrt{\mathrm{Hz}}$ ($0.9$ nV/$\sqrt{\mathrm{Hz}}$), currently limited by bias-line noise.  With sub-micrometer gaps and $Q>10^8$ membranes already demonstrated \cite{Tsaturyan2017,Ghamidi2018,Seis2022}, our model predicts that sub-200 fV/$\sqrt{\mathrm{Hz}}$ sensitivity is within reach, paving the way towards quantum-grade rf electrometers and heterodyne links that integrate seamlessly with microwave quantum circuits.

\section{Transduction principle}

\begin{figure}[h]
    \centering
    \includegraphics[width=0.95\linewidth]{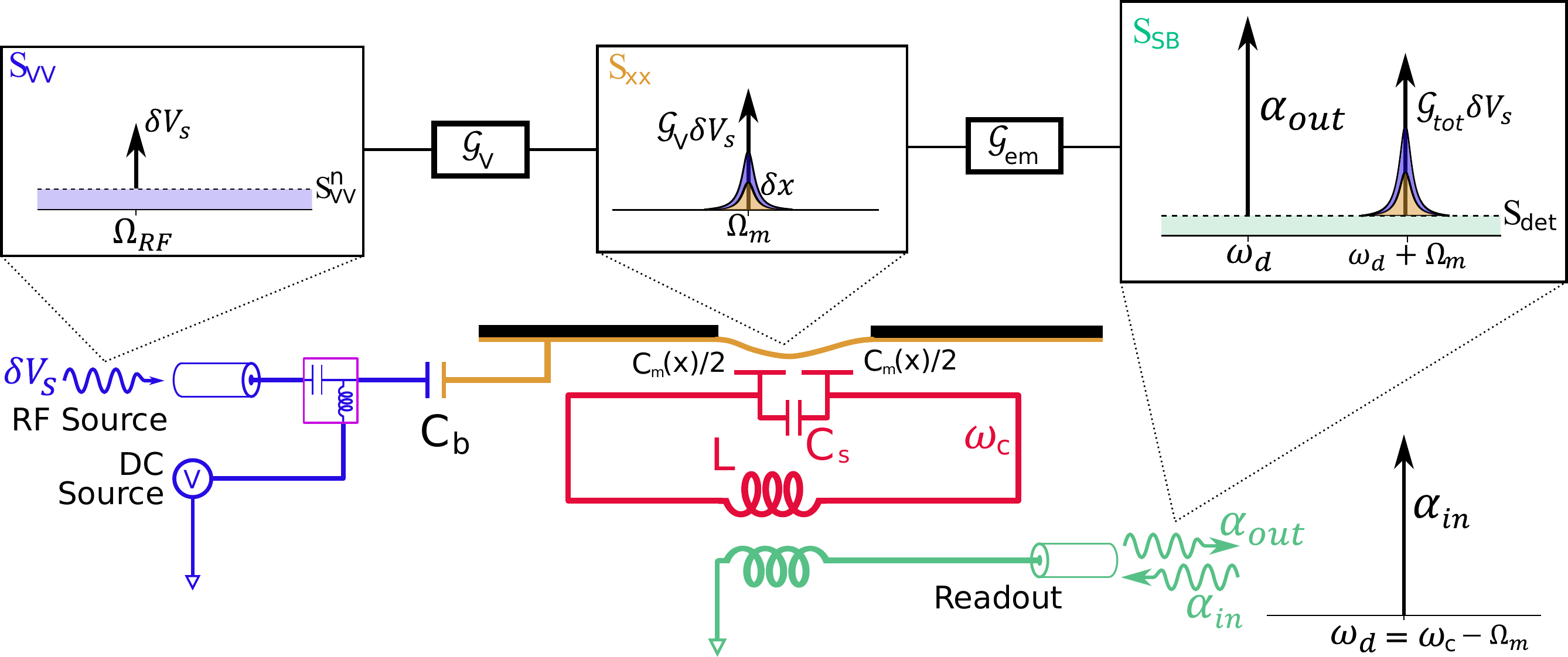}
    \caption{Top: illustration of the transducer amplification chain, where the electrical spectrum  $S_\mathrm{VV}$ (signal $\delta V_s$ at frequency $\Omega_\mathrm{RF}$ plus white noise of spectral density $S_\mathrm{VV}^n$, indicated in purple) undergoes transduction to position spectrum $S_\mathrm{xx}$  with the electrostatic gain $\mathcal{G}_\mathrm{V}$ (electrical noise contribution appears in purple, and thermal noise contribution in orange) followed by transduction to microwave sideband spectrum $S_\mathrm{SB}$ via the electromechanical gain $\mathcal{G}_\mathrm{em}$ (detection noise contribution is indicated in light green). Bottom: proposed implementation of the transducer. An LC microwave resonator (red) of resonance frequency $\omega_\mathrm{c}$ is coupled to a metalized free standing membrane acting as a mechanically compliant capacitor. The red part of the circuit is electrically floating. The membrane electrode (yellow) is biased through a capacitor $C_\mathrm{b}$, connected to a bias tee (pink rectangle) that combines the dc bias and an rf signal to drive membrane modes on resonance. The two halves of the membrane electrode behave as
\(C_m(x)/2\) capacitors that are connected in parallel with the microwave LC resonator, and in series with the bias-line, so that the bias circuit sees the full membrane capacitance \(C_m(x)\) whereas the LC resonator sees only \(C_m(x)/4\) in parallel with the stray capacitance \(C_\mathrm{s}\).
  The LC resonator is inductively coupled to a read-out line (green). A probe tone \(\alpha_\mathrm{in}\) at the red‑detuned frequency
\(\omega_\mathrm{d}=\omega_\mathrm{c}-\Omega_\mathrm{m}\) is sent into the
resonator, and the reflected field
\(\alpha_\mathrm{out}\) is monitored with a vector spectrum analyser. In this scheme, the mechanical mode of amplitude $\delta x$, is driven on resonance by the rf source, and amplified by the dc source with a gain $\mathcal{G}_V$. It is then upconverted and amplified with a gain $\mathcal{G}_\mathrm{em}$ at microwave frequencies by the electromechanical system operating in the resolved sideband regime. It appears as a sideband of the reflected microwave carrier $\alpha_\mathrm{out}$.}
    \label{fig:rf_transduction}
\end{figure}

The proposed design, described in Fig.~\ref{fig:rf_transduction}, consists of a metalized membrane oscillator (mode frequency $\Omega_\mathrm{m}/2\pi$ of a few MHz, and effective mass $m_\mathrm{eff}$), dispersively coupled to a superconducting lumped-element LC resonator (frequency $\omega_\mathrm{c}/2\pi$ of a few GHz). The capacitance of the LC resonator consists of a stray capacitance $C_\mathrm{s}$ in parallel with two mechanically compliant vacuum gap capacitance $C_\mathrm{m}(x)/2$ in series. The metalized membrane constitutes one plate of each capacitor $C_\mathrm{m}(x)/2$, such that its out-of-plane motion $x(t)$ shifts $\omega_\mathrm{c}$ according to $\omega_\mathrm{c}(x)
  = \left(L \left(C_\mathrm{m}(x)/4+C_\mathrm{s}\right)\right)^{-1/2}$.
To first order, this leads to a modulation of the microwave resonator frequency $\delta\omega_\mathrm{c}(t)=\left(\partial_x\omega_\mathrm{c}\right) x(t)$. The LC resonator is weakly coupled to a read-out-line and probed in reflection. A coherent microwave tone is applied with a frequency $\omega_\mathrm{d}$ typically red-detuned by one mechanical frequency: $\omega_\mathrm{d}\approx \omega_\mathrm{c}-\Omega_\mathrm{m}$.
In the linear regime, the reflected field acquires sidebands~\cite{Aspelmeyer2014} at $\omega_\mathrm{d}\pm\Omega_\mathrm{m}$, with a power spectral density expressed in $V^2/\mathrm{Hz}$
\begin{equation}
\label{Eq:sideband_spectrum}
  S_\mathrm{SB}[\omega]
  = \mathcal{G}_\mathrm{em}^2(\omega)
    S_\mathrm{xx}[\omega],
\end{equation}
where $S_\mathrm{xx}[\omega]$ is the power spectrum of the membrane position, and $\mathcal G_{\mathrm{em}}(\omega)$ the electromechanical gain, which can be approximated by 
\begin{equation}
 \mathcal G_{\mathrm{em}}(\omega)
  \simeq G\sqrt{n\kappa_c}\left|A_-(\omega)\right|\sqrt{\hbar\omega_\mathrm{c} Z/2}
  \end{equation}  
in the resolved sideband regime ($\Omega_\mathrm{m}\gg\kappa$), with $n$ being the microwave mode photon number, $G=\partial_x\omega_\mathrm{c}$
the electromechanical coupling strength,  
$A_-(\omega)=[-i(\Delta+\omega)+\kappa/2]^{-1}$
the microwave resonator response at detuning $\Delta=\omega_\mathrm{d}-\omega_\mathrm{c}$, and $Z$ is the read-out line characteristic impedance. Here, we have introduced the microwave resonator linewidth $\kappa=\kappa_c+\kappa_i$, the coupling rate to the read-out line $\kappa_c$, and the intrinsic loss rate of the microwave mode $\kappa_i$. This electromechanical system is used to upconvert and amplify the motion of the mechanical oscillator to the microwave domain. This technique is the standard read-out protocol for microwave optomechanical systems~\cite{Teufel2011, Seis2022}.

In addition, we add a static bias voltage $V$ and a small rf sinusoidal modulation $\delta V(t)$ on the membrane electrode through an external bias capacitor $C_\mathrm{b}$ (see Fig.~\ref{fig:rf_transduction}). Because $C_\mathrm{b}$ is in series with $C_\mathrm{m}$, the two capacitors can be replaced by the equivalent capacitance $C_{\mathrm{eq}}=\bigl(C_\mathrm{m}^{-1}+C_\mathrm{b}^{-1}\bigr)^{-1}.$ $V+\delta V(t)$ is applied across $C_{\mathrm{eq}}$, so that only the fraction $
\eta(V+\delta V(t))$, with 
$\eta=C_\mathrm{b}/(C_\mathrm{b}+C_\mathrm{m})$ drops across the mechanical capacitor $C_\mathrm{m}(x)$, the factor $\eta$ acting as a voltage dilution coefficient.
  The electrostatic potential energy of the system writes $
U(x,t)=-C_{\mathrm{eq}}(x) [V+\delta V(t)]^{2}/2.$
Expanding to first order in $\delta V/V$ yields the force
\begin{equation}
F_{\mathrm{el}}(t)\equiv-\partial_x U
   \simeq \frac{1}{2} V^{2}\partial_x C_{\mathrm{eq}}
     +V\delta V(t)\partial_x C_{\mathrm{eq}}.
\end{equation}
The first term is a static electrostatic force that both softens the membrane spring—producing the familiar anti‑spring frequency shift~\cite{Unterreithmeier2009, Bagci2014,Pate2018,Zhou2021,Gerashchenko2025} $\Omega_\mathrm{m}^2(V) = \Omega_\mathrm{m}^2(0)-V^2\partial_{x}^2C_\mathrm{eq}/2$, —and increases the mechanical linewidth through motion‑induced currents in any resistive elements $R$ of the biasing circuit, with the bias‑dependent damping $\Gamma_\mathrm{m}(V)=\Gamma_\mathrm{m}(0)+V^2(\partial_{x}^2C_\mathrm{eq})^2R/m_\mathrm{eff}$ (detailed calculations can be found in Ref.~\cite{Patange2025}).  The second term drives the membrane motion, with a force scaling linearly with $V$. The mechanical mode response to $F_\mathrm{el}(t)$ is given by an effective voltage-dependent  mechanical susceptibility $\chi_\mathrm{m}(\omega, V)$ defined by 
\begin{equation}
    \chi_\mathrm{m}(\omega, V)^{-1}=  m_\mathrm{eff}\left[\Omega_\mathrm{m}^{2}(V)-\omega^{2}-i\omega\Gamma_\mathrm{m}(V)\right]. 
\end{equation}  
Hence, an rf signal of spectrum $S_{\mathrm{VV}}[\omega]$ is transduced into displacement with the spectrum 
\begin{equation}
    S_{\mathrm{xx}}^{\mathrm{el}}[\omega]=\mathcal G_V(\omega)^2S_{\mathrm{VV}}[\omega], 
\end{equation}
where we have introduced the electrostatic gain
\begin{equation}
  \mathcal G_V(\omega)
  = \left|\chi_\mathrm{m}(\omega, V)V\partial_x C_\mathrm{eq}\right|,
\end{equation}
which grows linearly with $V$ and peaks at
$\omega=\Omega_{m}(V)$. The resulting overall transduction gain from an rf tone amplitude to a microwave sideband amplitude writes
\begin{equation}
\mathcal{G}_\text{tot}(\omega)=\mathcal{G}_{\text{em}}(\omega)\times \mathcal{G}_V(\omega).
\label{Eq:total_gain}
\end{equation}
The proposed device comprises two cascaded amplifiers: the electromechanical stage, whose gain \(\mathcal{G}_{\text{em}}\) depends on the microwave pump power and on the  electromechanical coupling $G$, and the electrostatic stage, whose gain \(\mathcal{G}_V\) can be tuned \emph{in situ} by the dc bias $V$. This shows that such transducer can be used to up-convert to the microwave domain and amplify small rf voltage fluctuations. In the following, we perform a noise budget, compute an expression for the total transduction gain, and derive an ultimate sensitivity of such a system used as an electrometer or charge sensor.

\section{Signal-to-noise ratio in an rf-to-microwave transduction experiment}

Force and position noise spectra are related by the mechanical susceptibility
\begin{equation}
    S_{\mathrm{xx}}[\omega] = \left|\chi_\mathrm{m}(\omega, V)\right|^2 S_{\mathrm{FF}}[\omega].
\end{equation}
Thermal force noise $S_{\mathrm{FF}}^{\mathrm{th}}$, electrical force noise $S_{\mathrm{FF}}^{\mathrm{el}}$ and cavity back-action force noise~\cite{Aspelmeyer2014} $S_{\mathrm{FF}}^{\mathrm{ba}}$ contribute to the total force noise as
\begin{equation}
     S_{\mathrm{FF}}[\omega] = S_{\mathrm{FF}}^{\mathrm{th}}[\omega] + S_{\mathrm{FF}}^{\mathrm{el}}[\omega] + S_{\mathrm{FF}}^{\mathrm{ba}}[\omega],
\label{eqn:Sxx_forced_drive}
\end{equation}
where
\begin{equation}
    S_{\mathrm{FF}}^{\mathrm{th}}[\omega] = 2m_\mathrm{eff}\Gamma_\mathrm{m} k_B T
    \label{eq:langevin}
\end{equation}
is the Langevin thermal force noise, with $T$ the environmental temperature, and $S_{\mathrm{FF}}^{\mathrm{el}}[\omega] = \left(\mathcal G_V[\omega]/|\chi_\mathrm{m}(\omega, V)|\right)^2 S_\mathrm{VV}[\omega]$ the force noise due to direct electrical drive on the membrane electrode. Including both  $S_{\mathrm{FF}}[\omega]$ and the detection noise $S_\mathrm{det}$, the transduced sideband spectrum writes
\begin{equation}
    S_{\mathrm{SB}}[\omega] 
    = \mathcal{G}^2_\mathrm{tot}(\omega)S_\mathrm{VV}(\omega) + 
    \mathcal{G}^2_\mathrm{em}(\omega)
    |\chi_\mathrm{m}(\omega, V)|^2 \left(S_\mathrm{FF}^\mathrm{th} + S_\mathrm{FF}^\mathrm{ba}\right) + S_\mathrm{det},\label{eqn:s_out_tot_temp_ch3}
\end{equation}
where the first term in the right-hand side is the term of interest. Splitting the input rf noise spectrum into signal and noise contributions as $S_\mathrm{VV} = S_\mathrm{VV}^\mathrm{s} + S_\mathrm{VV}^\mathrm{n}$, we introduce the signal to noise ratio of the transduction experiment:
\begin{equation}
\mathrm{SNR}^2 = \frac{\mathcal G_{\mathrm{tot}}^2(\omega) S_\mathrm{VV}^\mathrm{s}}{\mathcal{G}^2_\mathrm{tot}(\omega)S_\mathrm{VV}^\mathrm{n} + 
    \mathcal{G}^2_\mathrm{em}(\omega)
    |\chi_\mathrm{m}(\omega, V)|^2 \left(S_\mathrm{FF}^\mathrm{th} +  S_\mathrm{FF}^\mathrm{ba}\right) + S_\mathrm{det}}.
    \label{eqn:snr}
\end{equation}

Eq~\ref{eqn:snr} shows that the SNR can be improved by increasing the total transduction gain $\mathcal{G}_\mathrm{tot}$ as long as the dominant noise in the denominator is not amplified by the same factor. The total gain is controlled by two parameters: the electrostatic pre-amplification set by the dc bias $V$ and the microwave stage set by the intracavity photon number $n$ through the linearized coupling
$g_0\sqrt{n}$, with $g_0=Gx_\mathrm{zpf}$  the vacuum electromechanical coupling and $x_\mathrm{zpf}=\sqrt{\hbar/2m_\mathrm{eff}\Omega_\mathrm{m}}$ the zero point fluctuations of position. In the low-cooperativity limit this yields the
scaling $\mathcal{G}_\mathrm{tot}\propto |V|\sqrt{n}$. Therefore, in the imprecision-limited regime (denominator dominated by $S_\mathrm{det}$)
one expects $\mathrm{SNR}\propto |V|\sqrt{n}$. By contrast, when the transduced input electrical noise term
$\mathcal{G}_\mathrm{tot}^2S_\mathrm{VV}^\mathrm{n}$ dominates, increasing $\mathcal{G}_\mathrm{tot}$ amplifies signal
and noise identically and the SNR saturates towards $\sqrt{S_\mathrm{VV}^\mathrm{s}/S_\mathrm{VV}^\mathrm{n}}$.
Finally, at high pump power, the backaction contribution increases (with $S_\mathrm{FF}^\mathrm{ba}\propto n$), so that the SNR
generally exhibits an optimum as a function of $n$~\cite{Aspelmeyer2014,Zhou2021APL}.

\section{Total transduction gain and ultimate sensitivity of the transducer}

To illustrate the potential of such a transducer, we compute the total gain and the thermal noise limited sensitivity in the low cooperativity limit $C =  4g_0^2n/\kappa\Gamma_\mathrm{m}\ll 1$, where cavity back-action force noise can be neglected. The total transduction gain at detuning $\Delta=-\Omega_\mathrm{m}$ is 
\begin{equation}
 \mathcal{G}_\mathrm{tot}(\Omega_{m}) = \sqrt{\frac{C\kappa_c Z\omega_\mathrm{c}}{\kappa\Omega_\mathrm{m} m_\mathrm{eff}\Gamma_\mathrm{m}}}\left|V\partial_x C_\mathrm{eq}\right|.
 \end{equation}
Additionally, the ultimate sensitivity is obtained for an SNR of 1. Assuming that the detection noise $S_\mathrm{det}$ and the electrical input noise $S_\mathrm{VV}^\mathrm{n}$ can technically be made negligible, we get
\begin{equation}
 \sqrt{S_{\mathrm{VV}}^{\mathrm{min}}} = \frac{\sqrt{2m_{\mathrm{eff}}\Gamma_{m} k_B T}}{\left|V\partial_x C_\mathrm{eq}\right|}.
\end{equation}

Let us estimate the gain and sensitivity that could be obtained with state-of-the art sub-systems. To date, highly stressed silicon nitride membrane resonators are the best candidates to perform such an experiment. They can have frequencies around 4~MHz, masses around 1~ng, and they can be thermalized at a temperature of 10~mK in a dilution cryostat~\cite{Gerashchenko2025}, exhibiting quality factors of at least 100 million, which leads to $\Gamma_\mathrm{m}/2\pi$ smaller than 10~mHz. Such extreme quality factors can be obtained through phononic engineering in silicon-nitride stressed structures~\cite{Tsaturyan2017, Yu2014, Cupertino2024, Bereyhi2022, Seis2022, Patange2025}, relying on dissipation dilution~\cite{Fedorov2019}. Finally, sub-micrometer distances can be reached in vacuum gap capacitors~\cite{Seis2022}. For a 500~nm distance and a surface of 20~$\mu$m$\times$20~$\mu$m, we find $C_\mathrm{m}\simeq 10~$fF, and  $\partial_x C_\mathrm{m} \approx 15~$nF/m. We assume that $C_\mathrm{b}$ is large enough such that $\eta\simeq 1$. We also assume a maximum dc bias of \(50\)~V: while cryostat dc lines and electronic components can typically handle up to \(150\)~V, avalanche breakdown in semiconducting substrates usually imposes a lower practical limit~\cite{Sze2012, Gerashchenko2025}.
In this configuration, assuming a cooperativity of $C = 0.1,~\kappa_c/2\pi=0.5$~MHz, and $\kappa_i/2\pi=1$~MHz, the total gain would be $\mathcal{G}_\mathrm{tot}(\Omega_\mathrm{m})=44$~dB, and the  minimum voltage modulation that could be detected is of the order of 200~fV$/\sqrt{\mathrm{Hz}}$, or 10~ne/$\sqrt{\mathrm{Hz}}$ in terms of charge modulation, rivaling with the lowest sensitivities reported so far~\cite{Cassidy2007, Viennot2014, Gonzalez2015, Najera2024}.

\section{Experimental implementation}

In this section, we demonstrate an experimental implementation of such an rf-to-microwave transducer.

\begin{figure}[h]
    \centering
    \includegraphics[width=0.94\linewidth]{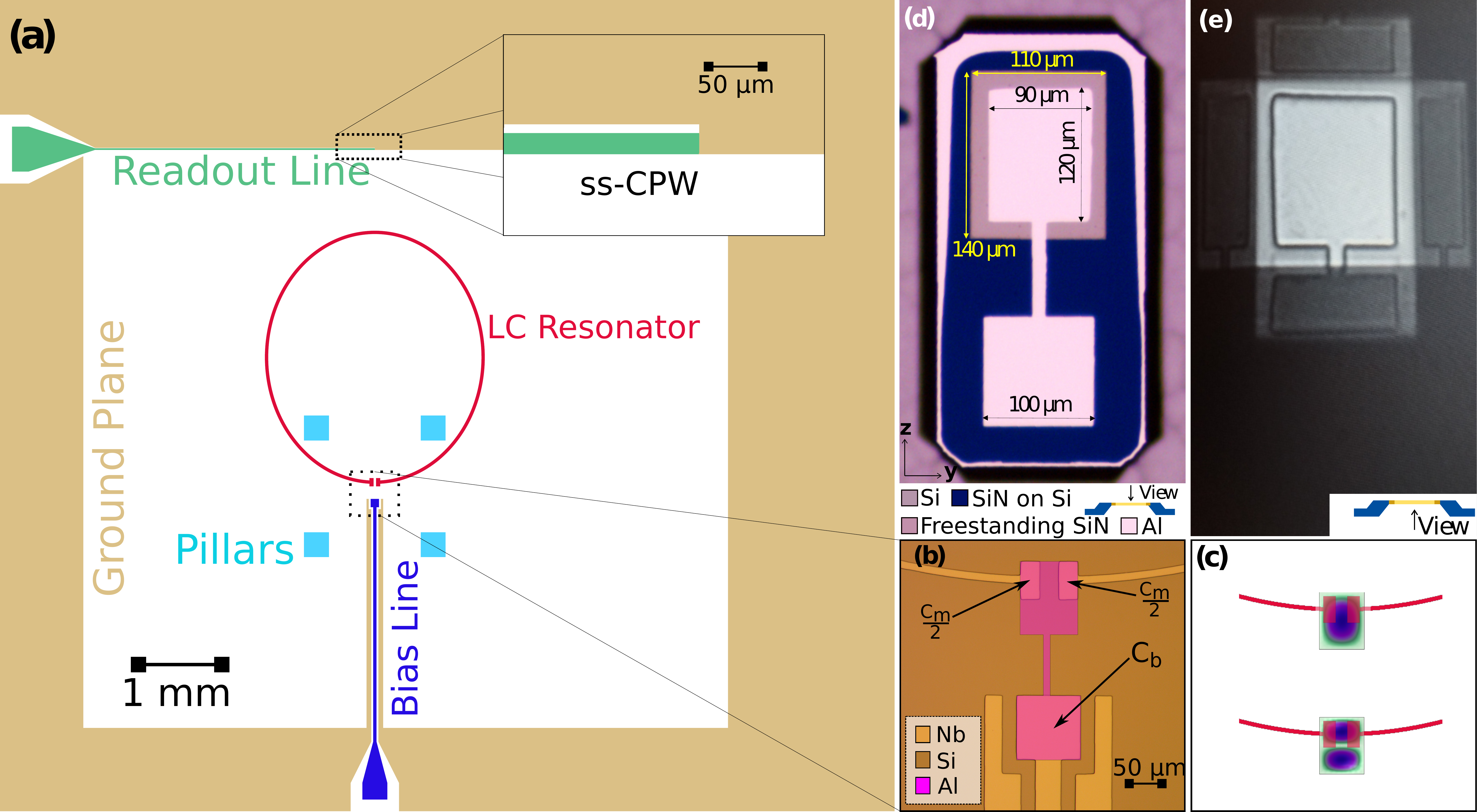}
    \caption{(a) LC resonator chip design: LC resonator (red), biasing line (blue), the read-out line (green), the ground plane (gold), and the aluminum pillars (light blue) used as spacers in the flip-chip process. For visualization, the inductor track (width: 8 \textmu m) has been thickened. \emph{Inset:} close-up on the read-out line which is a single-sided coplanar waveguide terminating into the ground plane. (b) Optical microscope image of the capacitor pads of the LC resonator, overlaid with the membrane electrodes (pink). (c) The LC resonator (red) overlaid with simulated mode profiles of the first to membrane modes 11 and 12. (d) Optical microscope image of the membrane sample, standing on a rectangular pedestal 25~$\mu$m above the silicon substrate. The pedestal with the membrane and electrodes is in focus, the substrate is not in focus. (e) Optical microscope image of the membrane sample, seen from the other side. The released metalized SiN membrane can be seen at the top. Mirror images on the four edges of the membrane come from the smooth inclined silicon crystalline planes obtained after etching with KOH trough the substrate. (d) and (e) figures are taken from Ref.~\cite{Gerashchenko2025}.}
    \label{fig:LC_and_memb}
\end{figure}

\subsection{Membrane oscillator}

The mechanical oscillator is a 90~nm-thick, high-stress ($\sigma\simeq$1~GPa) Silicon Nitride (SiN) rectangular membrane with dimensions $l_y=$110~\textmu m, and $l_z=$140~\textmu m. The combination of high tensile stress and low thickness yields high quality factors thanks to dissipation dilution~\cite{Fedorov2019}. The central part of the membrane (rectangle of dimensions 90~\textmu m$~\times$ 120~\textmu m) is coated with a 30~nm-thick aluminum layer, leaving a narrow uncoated stripline at the edges. The aluminum biasing circuit consists of a square aluminum pad of 100~\textmu m side length, forming one electrode of $C_\mathrm{b}$, that is electrically connected to the membrane pad with a thin stripline. A schematic of this layout is shown in Fig.~\ref{fig:LC_and_memb}.

The fabrication process is the following. A silicon wafer is coated on both sides with a (LPCVD) 100~nm SiN layer, and is annealed at \SI{1100}{\celsius}~\cite{Mittal2024} for four hours. The membranes are released by wet etching at \SI{85}{\celsius} in a  30~\% KOH solution, after performing an optical lithography step~\cite{Ivanov2020, Ivanov2020data, Gerashchenko2025} on one side of the wafer. Additionally, a second optical lithography step followed by KOH etching on the other side of the wafer (the membrane side) defines a 25~µm‑deep recess over the entire chip except the membrane and dc bias circuit footprint (Fig.~\ref{fig:LC_and_memb}(d)). The resulting relief maintains a generous inter‑chip spacing everywhere outside the active membrane region, so that in the final assembly, any dust particles trapped there cannot affect the capacitor gap beneath the membrane. The membranes are dipped in a 10\% diluted HF solution to remove any dirt and debris. During this step, HF etches 10~nm of SiN in total, leaving a 90~nm-thick membrane. The dc biasing electrodes are fabricated out of aluminum using a lift-off process.

The mechanical modes frequencies, profiles, and quality factors were characterized at room temperature in an optical interferometer prior to metalization, and show frequencies of 3.22~MHz (mode 11) and 4.73~MHz (mode 12), in good agreement with the result expected from the plate theory~\cite{leissa_vibration_1969} $\Omega_\mathrm{pq} = \pi\sqrt{(\sigma/\rho)\left(p^2/l_\mathrm{y}^2 + q^2/l_\mathrm{z}^2\right)}$, with $p$ and $q$ being the integer mode indices, $\rho$ the membrane density. Finally, mechanical quality factors were measured near $6\times10^4$ prior to aluminum metalization.

\subsection{LC resonator}

The LC resonator capacitor is made of two rectangular pads (size 60~\textmu m $\times$ 90~\textmu m) designed such that the microwave mode couples primarily to the first two membrane modes (modes 11 and 12 on Fig.~\ref{fig:LC_and_memb}(c)). They are connected to a nearly circular ring inductor. The LC resonator chip also hosts a 100~\textmu m$\times$ 100~\textmu m metal pad forming one electrode of $C_\mathrm{b}$. Fig.~\ref{fig:LC_and_memb}(a, b) show the biasing pad and the biasing line (blue in (a)), the membrane electrode, the inductor, and their alignment with the membrane chip. On the side opposite to the capacitor electrodes, the inductor couples to a single-sided coplanar waveguide read-out line, impedance-matched to 50~$\Omega$ (see Fig.~\ref{fig:LC_and_memb}(a) and inset). Finite‑element simulations with Ansys HFSS are used to optimize the circuit geometry. A gap‑to‑track ratio of $4/15$ yields a 50~$\Omega$ characteristic impedance, matching that of the external line. The simulations are also used to set the fundamental resonance at $\omega_\mathrm{c}/2\pi = 7$~GHz and to adjust the mutual inductance between the coplanar waveguide and the LC resonator. The superconducting circuit is fabricated by sputtering a niobium film onto an intrinsic silicon wafer, then patterning the film with optical lithography, followed by reactive‑ion etching.
\begin{figure}[h!]
    \centering
    \includegraphics[width=0.95\linewidth]{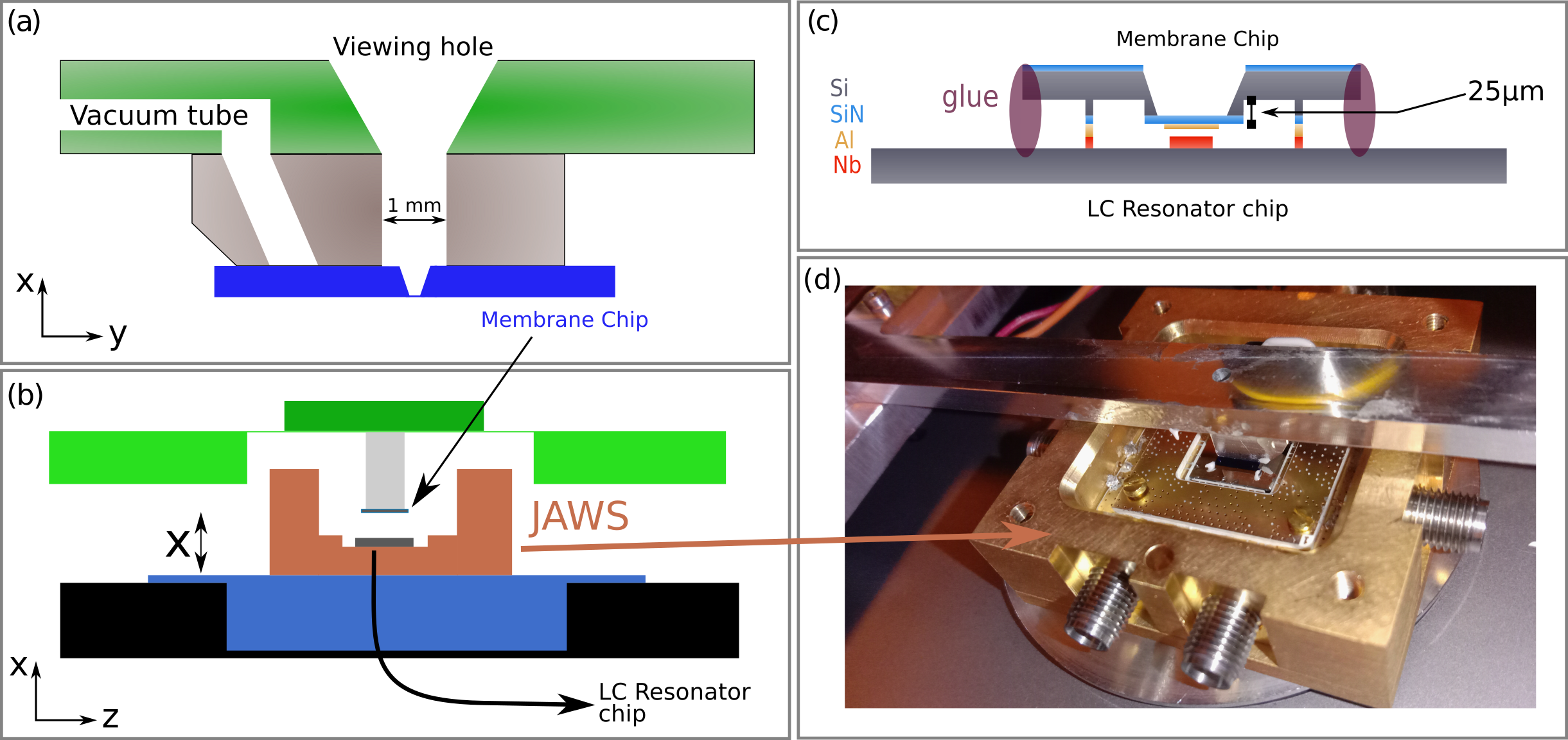}
    \caption{(a) Illustration of the custom mask holder used in the flip-chip process, showing the membrane chip (blue) being picked up via a vacuum suction tube. A viewing hole allows visual alignment. (b) We use the 3 axis micropositioning of the MJB4 mask aligner system under binocular microscope to align the membrane chip with the LC resonator chip that is already wirebonded in a microwave-shielding box, and bring them in contact. (c) Cross-sectional schematic of the final device. The membrane pedestal at a distance of 25~\textmu m from the substrate can be seen in the center. On the LC resonator chip, the 600~nm-tall aluminum pillars (not to scale) are used as spacers to set the inter-chip distance. Glue drops are also shown at the chip edges. (d) Photograph of the completed flip-chip assembly in the mask aligner, with chips glued together (white drops indicate adhesive). The microwave-shielding box is indicated.}
    \label{fig:flipchip_process}
\end{figure}

\subsection{Flip-chip assembly}

Capacitive coupling between the membrane and the LC resonator, is obtained via a flip-chip assembly technique~\cite{Capelle2020, Patange2025}. The membrane and circuit electrodes are brought in close proximity, forming the vacuum gap capacitors $C_\mathrm{b}$ (bias-to-membrane) and $C_\mathrm{m}$ (membrane-to-LC circuit). The chip distance is set by four 600-nm aluminum spacers, evaporated on the LC resonator chip.
The assembly is performed on an MJB4 mask aligner using a custom-designed mask holder, used to pick up the membrane chip and align it on top of the LC resonator chip (see Fig.~\ref{fig:flipchip_process}). Prior to alignment, the LC resonator chip has been glued inside a microwave shielding box  (Joint Assembly for the Wiring of Superconducting circuits~\cite{Villiers2023}), in which the various required rf, microwave and dc ports are wirebonded to macroscopic connectors (see Fig.~\ref{fig:flipchip_process}). A viewing hole in the holder provides top-down visibility of the membrane. Once aligned, the two chips are glued together in place using Dymax OP-67-LC UV-curing adhesive.

The LC resonator frequency is measured at $\omega_\mathrm{c}/2\pi = 7.04~$GHz before flip-chip and $\omega_\mathrm{c}/2\pi = 6.21~$GHz after flip-chip, from which we estimate the participation ratio of the vacuum-gap capacitor $C_\mathrm{m}/(C_\mathrm{m} + 4C_\mathrm{s}) \sim 0.22$. The measured linewidths are  respectively $\kappa/2\pi \approx 1.6$~MHz and $\kappa/2\pi \approx 1.34$~MHz before and after the flip-chip process, ensuring that the system operates in the resolved sideband regime.

\subsection{Electromechanical Characterization of Mechanical Modes}
\begin{figure}[h!]
    \centering
    \includegraphics[width=1\linewidth]{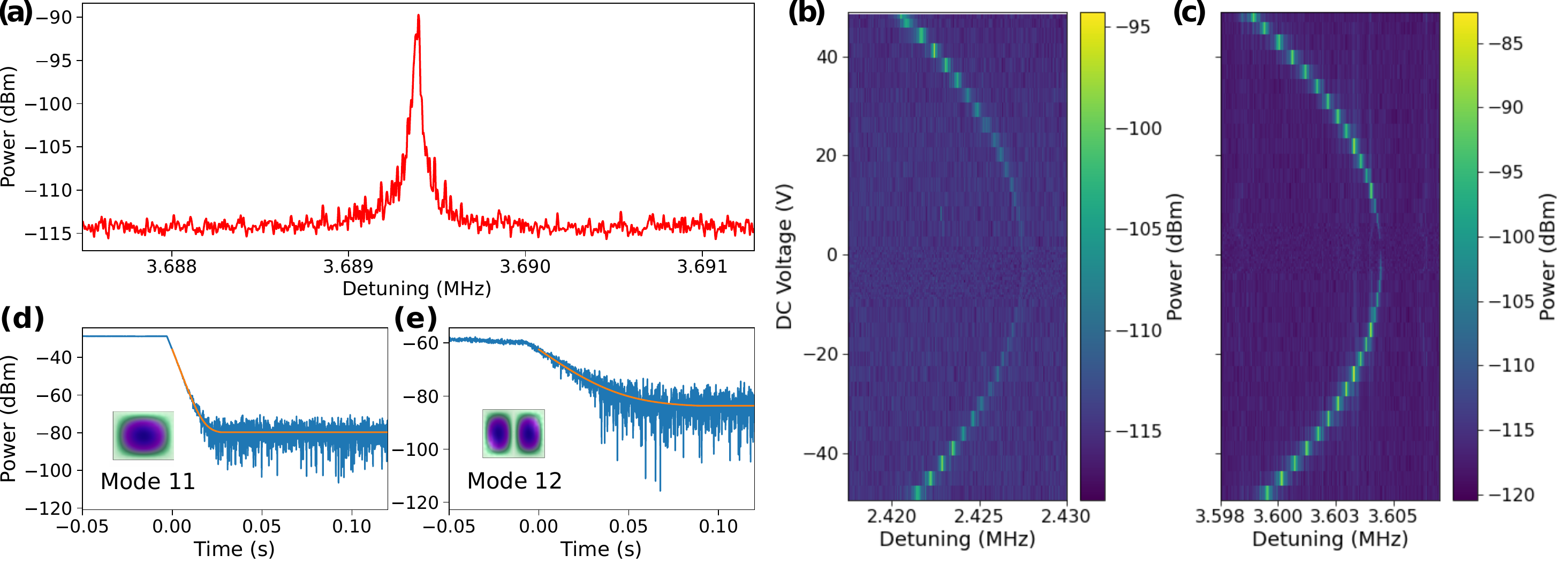}
    \caption{ (a) Sideband noise amplitude from mode 12, on the read-out reflected signal, with the microwave pump at $\omega_\mathrm{d} = \omega_\mathrm{c} - \Omega_\mathrm{m}$, and with a non zero dc bias $V=30~$V.
    (b) and (c) show the frequency shift of the two mechanical modes for $V\in [-49, 49]$~V, highlighting the anti-spring effect. (d) and (e): ring-down measurements for mode 11 and 12  performed at 10~mK near zero dc bias. The membrane mode is driven by a strong resonant rf tone through the biasing line, and its exponentially decaying amplitude  after turning off the rf tone, is monitored versus time. By fitting the decay with an exponential curve (yellow line) we extract $\Gamma_\mathrm{m}$.}
    \label{fig:mechanical_detections}
\end{figure}

The mechanical mode is probed by sending a microwave tone at $\omega_\mathrm{d} = \omega_\mathrm{c} - \Omega_\mathrm{m}^{pq}$ on the read-out port ($pq$ stand for the mechanical mode indices: 11 and 12), and monitoring the mechanical sideband on the reflected signal with a vector signal analyzer (VSA). A comprehensive cryostat wiring schematic can be found in the Supplementary Material.  An example of a VSA trace is shown in Fig.~\ref{fig:mechanical_detections}(a). Sideband spectra are plotted versus $V$, in Fig.~\ref{fig:mechanical_detections}(b), (c) for the two mechanical modes. We observe the expected anti-spring effect, and a linewidth increase by a factor $\lesssim 2$ between 0 and 49~V bias, which we attribute to heating via Joule effect. Analyzing this parabolic dependence reveals a membrane-LC resonator distance near 1.5~\textmu m, and zero bias frequencies of 2.4~MHz and 3.6~MHz for the modes 11 and 12 respectively ---the values having been shifted due to the added mass of aluminum. In Fig.~\ref{fig:mechanical_detections}(d) and (e), we show the mechanical mode amplitude versus time after resonant drive excitation through the rf port. Fitting an exponential decay to the data yields the decay rates $\Gamma^{11}_\mathrm{m}/2\pi=89$~Hz and $\Gamma_\mathrm{m}^{12}/2\pi = 16~$Hz. Note however, that in experiments performed in this work, a microwave pump power of -30~dBm results in a small cold damping effect~\cite{Aspelmeyer2014}. The measured modified decay rates are $\Gamma_\mathrm{em}^{11}/2\pi = 114~$Hz and $\Gamma_\mathrm{em}^{12}/2\pi = 24~$Hz. Finally, we use a standard calibration method~\cite{Gorodetksy2010} to extract $g_0^{11}/2\pi = 0.29\pm 0.07$~Hz and $g_0^{12}/2\pi = 0.23\pm 0.05$~Hz. At -30~dBm microwave pump power, chosen as the highest pump power compatible with stable cryostat temperature, it yields cooperativities of 0.28 and 0.5 for the modes 11 and 12, respectively, and provides the ratio between the two modes effective masses. To obtain an absolute value of the effective masses, we proceed in two steps. First, for a stress‑dominated rectangular membrane the motional mass is one quarter of the physical mass, a result that is independent of the mode indices~\cite{Tsaturyan2017}. Second, only the electrode area contributes to the
signal, such that the motional mass must be multiplied by a read‑out dilution factor, equal to the ratio of the squared mode amplitude integrated over the electrodes area, to that over the entire membrane. Defining the displacement coordinate as the maximum transverse amplitude, we find $m_\mathrm{eff}^{11} = 6.25~$ng and $m_\mathrm{eff}^{12} = 4.27~$ng, which is compatible with the measured mass ratio.
\begin{figure}[h!]
    \centering
    \includegraphics[width=0.75\linewidth]{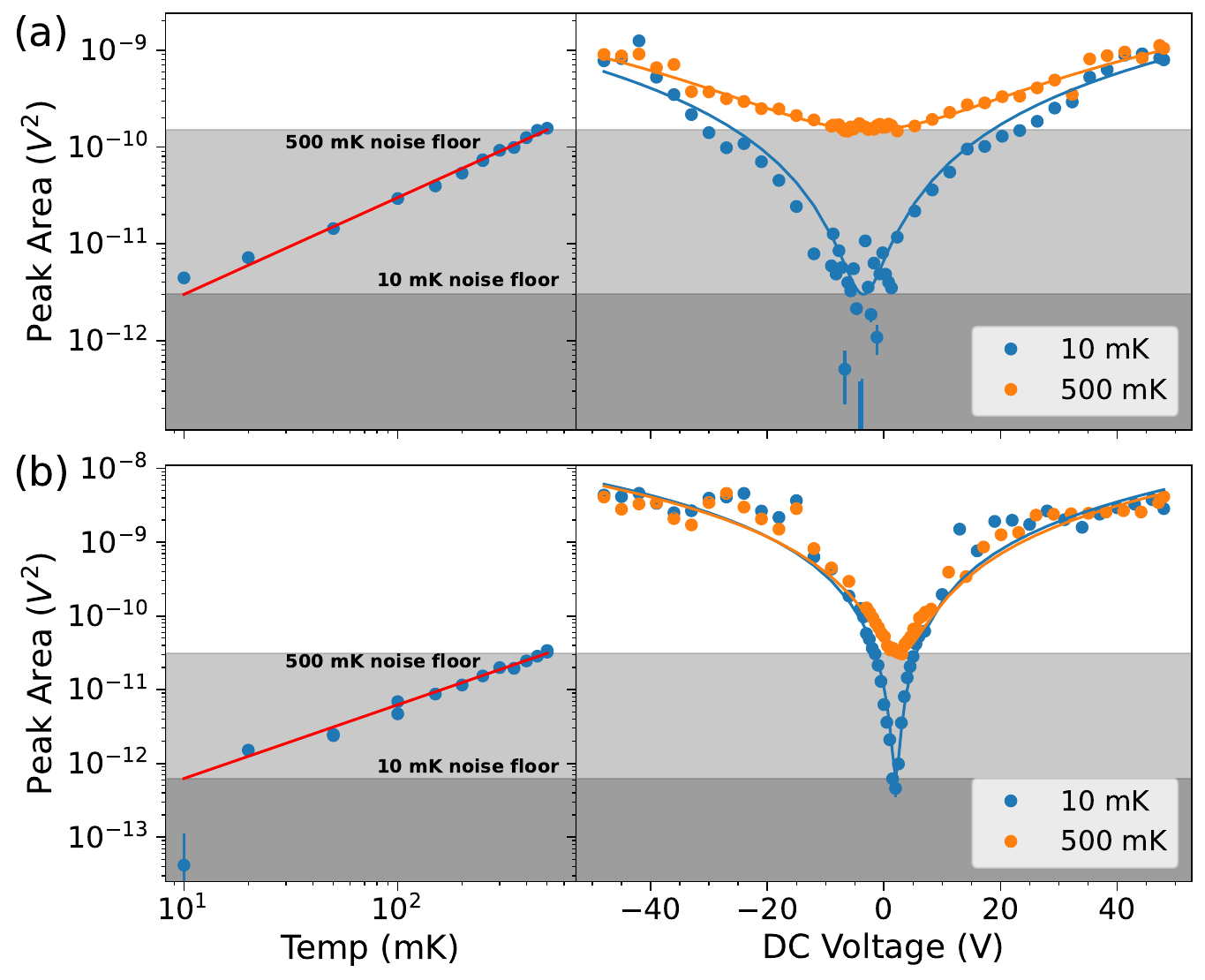}
    \caption{Membrane modes thermalization and electrical noise transduction.  Left column: transduced sideband spectrum area versus cryostat temperature at zero bias voltage. (a): mode 11. (b): mode 12. From a linear fit (red), we extract a slope of $3.57\times10^{-10}$~V$^2$/K for mode 11, shown in (a), and $9.8\times10^{-11}$~V$^2$/K for mode 12, after correcting for power broadening.
    The 10~mK area value and 500~mK area value of each mode defines the corresponding thermal noise floor highlighted in two different gray colors. Right column: area versus $V$ at two different temperatures. Eq.~\ref{eqn:area_v_mode} is fitted to the data (solid lines).}
    \label{fig:temp_voltage}
\end{figure}

\subsection{Mechanical modes thermalization and electrical noise transduction}
\label{sec:noise_budget}

The mechanical sideband area can be obtained by integrating Eq.~\ref{Eq:sideband_spectrum}. We find
\begin{equation}
\label{eqn:area_v_mode}
    \mathcal{A}_V = \frac{1}{{m_\mathrm{eff}\Omega^2_{m}(V)}}\left(k_BT\frac{\Gamma_\mathrm{m}}{\Gamma_\mathrm{em}} + \frac{S_{VV}^n}{m_\mathrm{eff}\Gamma_\mathrm{em}}\left(\partial_x C_\mathrm{eq}\right)^2 (V - V_0)^2 \right).
\end{equation}
The first term in the right-hand side gives the standard linear dependence with temperature. The second term in the rhs, is the parabolic dependence of the mechanical energy due to transduced electrical noise. Here the offset $V_0$ accounts for any trapped charge on the floating superconducting island. Scanning the cryostat temperature from 10~mK to 500~mK when the effective bias voltage is set to zero reveals the linear dependence of the peak area with temperature as shown in Fig.~\ref{fig:temp_voltage} (left column) for both modes, confirming that the membrane modes are thermalized to the cryostat base-temperature. Additionally, in Fig.~\ref{fig:temp_voltage} (right column), we plot the sideband area versus bias voltage. Eq.~\ref{eqn:area_v_mode} is fitted to the data  obtained at 10~mK and 500~mK, emphasizing the validity of our model. 

\subsection{SNR of transduction experiment}

\begin{figure}[h!]
    \centering \includegraphics[width=0.95\linewidth]{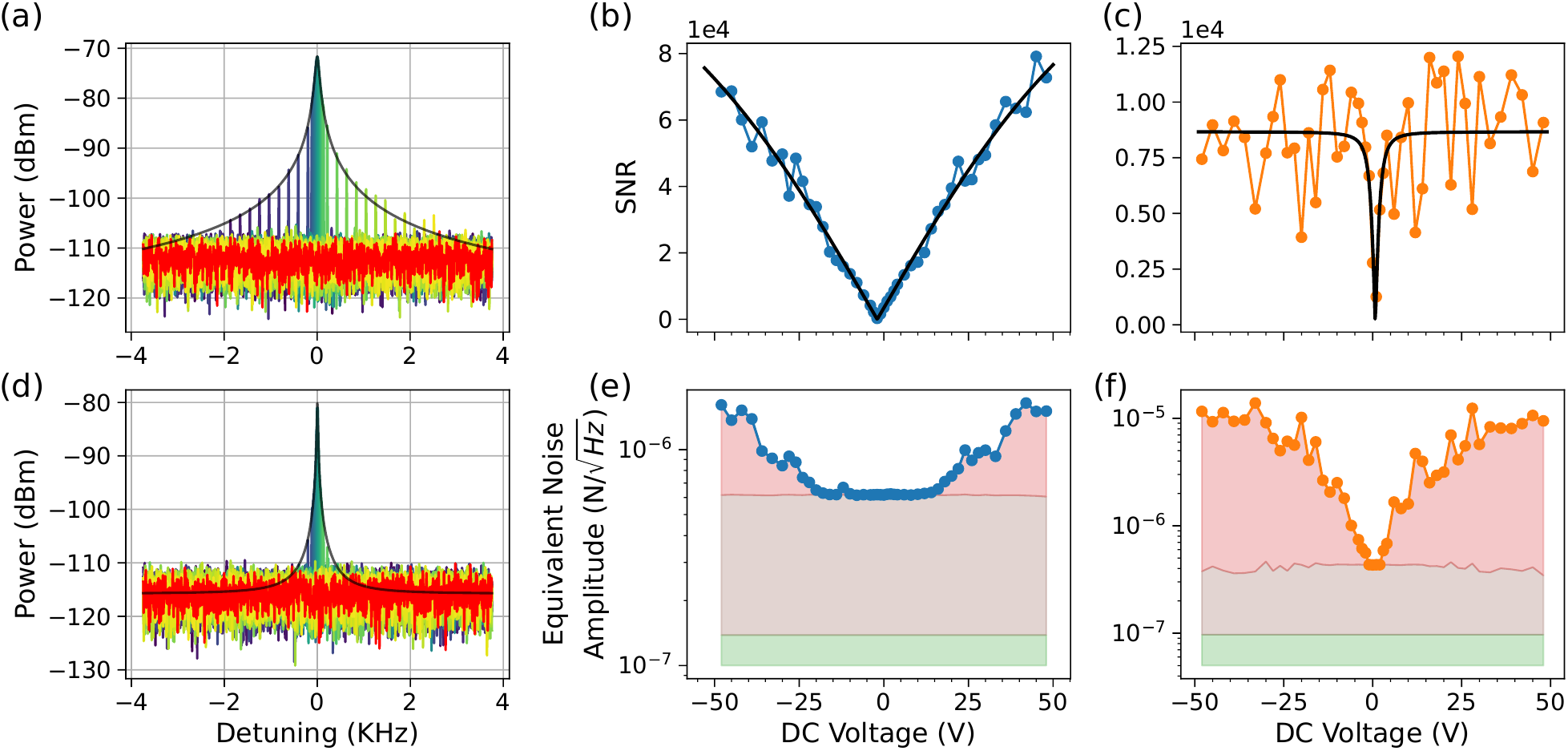}
    \caption{Rf signal transduction experiment for mode 11 in (a) and mode 12 in (d). An rf tone is  applied on the bias-line, and its frequency is swept through the mechanical mode resonance, while maintaining $V$ close to zero. The LC Resonator is probed via the read-out line like in previous experiments. The VSA traces, taken at various detuning $(\Omega_\mathrm{rf}-\Omega_\mathrm{m})/2\pi$, are shown with a gradient of colors from blue to green. The transduction envelope reveals the membrane mode susceptibility, and can be fitted with a lorentzian curve (black solid line). The red curve shows the VSA trace in absence of rf signal. (b) and (e) show the measured Signal to Noise Ratio (SNR) of the rf peaks on resonance, transduced by the membrane modes 11 and 12 respectively, with the dc bias varying from -49~V to 49~V, and with an rf drive of power -70~dBm. (c) and (f) respectively highlight the overall force noise budget: the gray filled area represent the detector noise, green filled area the thermal noise, and the red filled area shows the transduced electrical noise, in the data of (b) and (e).}
    \label{fig:transduction_experiment}
\end{figure}

At a non zero bias voltage, we sweep the frequency $\Omega_\mathrm{rf}$ of a sinusoidal rf drive across the mechanical mode frequency. The sideband spectra are shown in Fig.~\ref{fig:transduction_experiment} (a) and (d) at various  detuning  $\Omega_\mathrm{rf}-\Omega_\mathrm{m}$. As predicted by Eq.~\ref{eqn:s_out_tot_temp_ch3}, we observe an overall Lorentzian envelope stemming from the mechanical susceptibility. The noise spectra obtained in the absence of rf signal appears in red in Fig.~\ref{fig:transduction_experiment} (a) and (d). By monitoring the height of the transduced sideband spectrum on resonance, when $\Omega_\mathrm{rf} = \Omega_\mathrm{m}$, across the voltage range of -49~V to 49~V, and after normalizing with the resolution bandwidth of the VSA,  we extract the signal to noise ratio of the transduction experiment, as defined by Eq.~\ref{eqn:snr}. The results are shown in Fig.~\ref{fig:transduction_experiment} (b) and (e). The data reveal a linear dependence at small $V$, followed by a saturation, when entering in the regime where the transduced electrical noise dominates other noise contributions (detection and thermal noises). For mode 11, only the linear regime is accessible. For mode 12, the lower values of thermal noise and detector noise yields a factor $\sim 2$ steeper slope in the linear regime, sufficient to reach a saturation due to electrical noise.

To analyze the data we first rewrite Eq.~\ref{eqn:snr}  
\begin{equation}
\mathrm{SNR}^2 = \frac{ \left|V\partial_x C_\mathrm{eq}\right|^2 S_\mathrm{VV}^\mathrm{s}}{\left|V\partial_x C_\mathrm{eq}\right|^2S_\mathrm{VV}^\mathrm{n} + 
     S_\mathrm{FF}^\mathrm{th} +  S_\mathrm{FF}^\mathrm{ba} + S_\mathrm{FF}^\mathrm{det}},
    \label{eqn:snr2}
\end{equation}
where we have defined the equivalent force noise associated to detector noise $S_\mathrm{FF}^\mathrm{det} = S_\mathrm{det}/\allowbreak\left(\mathcal{G}^2_\mathrm{em}(\omega)
|\chi_\mathrm{m}(\omega, V)|^2\right)$. The term $S_\mathrm{FF}^\mathrm{th}+S_\mathrm{FF}^\mathrm{ba}$ is computed from Eq.~\ref{eq:langevin}, after substituting  $\Gamma_\mathrm{m}$ with $\Gamma_\mathrm{em}$ to account for backaction. $S_\mathrm{FF}^\mathrm{det}$ is then obtained experimentally from the noise spectra measured at zero effective dc voltage (spectra used to compute the area points in the left column of Fig.~\ref{fig:temp_voltage}). More precisely, we compute the ratio $(S_\mathrm{FF}^\mathrm{th}+S_\mathrm{FF}^\mathrm{det}+S_\mathrm{FF}^\mathrm{ba})/S_\mathrm{FF}^\mathrm{det}$ between the noise peak maximum and the noise floor away from the resonance. Using the computed value of $S_\mathrm{FF}^\mathrm{th}+S_\mathrm{FF}^\mathrm{ba}$ then yields $S_\mathrm{FF}^\mathrm{det}$. 
Next, we evaluate $\left|\partial_x C_\mathrm{eq}\right|$ by using geometrical parameters. Finally,  fitting Eq.~\ref{eqn:snr2} to the data, provides the values of $S_\mathrm{VV}^s[\Omega_\mathrm{m}]$ and $S_\mathrm{VV}^n$, which is assumed locally frequency-independent. We find $S_\mathrm{VV}^s[\Omega_\mathrm{m}] = -70.4$~dBm for mode 11 and $-70.3~$dBm for mode 12, in good agreement with the rf drive power of -70~dBm---estimated from the source power and the total attenuation of the rf line---confirming the distance value obtained from anti-spring effect. The value of $S_\mathrm{VV}^n (\Omega_\mathrm{m})$ is calculated only in the case of mode 12, for which the SNR saturation is reached. We find $\sqrt{S_\mathrm{VV}^n (\Omega_\mathrm{m}^{12})} = 7.8~\mathrm{nV/}\sqrt{\mathrm{Hz}}.$      Finally, in Fig.~\ref{fig:transduction_experiment}(e) and (f), we show the various noise contributions versus dc voltage: thermal and cavity backaction noise, detection noise, transduced electrical noise. Note, that in our present setup the cooperativity remains $C\lesssim 1$, and the detected background is dominated by amplification-chain added noise downstream of the device. Since this contribution does not enter the cavity, it does not generate imprecision–backaction correlations; we therefore neglect correlation terms in the analysis. Note, that if the dominant noise were incident on the cavity field (e.g., pump technical noise), correlations could impact the signal~\cite{Safavi-Naeini2013, Sudhir2017, Zhou2021APL, Seis2022}.

\subsection{Electrometer Sensitivity}
To evaluate the sensitivity of this electrometer, we determine the signal at which the SNR at maximum dc bias reaches 1. 
In practice, it is obtained by dividing the signal strength $S_\mathrm{VV}^\mathrm{s}$ by the highest measured SNR (value taken from the fit). The voltage and charge sensitivities for mode 11 and 12 are listed in Table~\ref{Table:sensitivity_values}.
\begin{table}[h]
\centering

\begin{tabular}{|c|c|c|c|c|c|c|}
\hline
\multirow{2}{*}{Mode} & \multirow{2}{*}{\makecell{Max SNR}}  & \multicolumn{2}{c|}{Measured Sensitivity} & \multicolumn{2}{c|}{Ideal Sensitivity} & \multirow{2}{*}{$\Gamma_\mathrm{m}/2\pi$ (Hz)}   \\
\cline{3-6}
& &  $\mathrm{nV/\sqrt{Hz}}$ &   \textmu$\mathrm{e/\sqrt{Hz}}$ &  $\mathrm{nV/\sqrt{Hz}}$ &   \textmu$\mathrm{e/\sqrt{Hz}}$ & \\
\hline
1  & 96~dB & 0.9 & 87 & 0.1 & 9 & 89 \\
2  & 78~dB & 7.8 & 760 & 0.05 & 5 & 16 \\
\hline
\end{tabular}
\vspace{0.1cm}
\caption{Maximum SNR, measured and ideal sensitivity values for the two modes. The values are obtained at -30~dBm microwave pump power, and -70~dBm RF tone power. The values in elementary charge per root Hertz are calculated from the biasing circuit capacitance $C_\mathrm{eq} \approx 15$~fF. The ideal sensitivity is the thermal noise limited sensitivity. All sensitivity values and maximum SNR are calculated/measured at 49~V, and $\Gamma_\mathrm{m}$ values are recalled in the last column.}
\label{Table:sensitivity_values}
\end{table}
Note that the sensitivity of mode 11 is limited by the maximal dc voltage value that can be reached in the experiment, whereas the sensitivity of mode 12 is limited by electrical noise. Our interpretation is that electrical noise might be significantly different at the two mode frequencies. 

\section{Conclusion}\label{sec:conclusion}

We have presented a dc-biased membrane-in-LC architecture that performs heterodyne conversion of radio-frequency voltages to the microwave band, while acting as a sensitive electrometer. By decoupling the total gain from microwave pump power through electrostatic pre-amplification, we avoid the excess shot noise and heating that constrain conventional cavity converters, at the cost of adding electrical noise, which can however be mitigated by better filtering strategies, or using symmetry protection~\cite{Gerashchenko2025}. In a flip-chip device with a 1.5~\textmu m vacuum gap, we observe the expected anti-spring tuning effect on the mechanical modes, and achieve a charge sensitivity of 87~\textmu e$/\sqrt{\mathrm{Hz}}$ (0.9 nV$/\sqrt{\mathrm{Hz}}$). By implementing a related concept at 10~mK (rather than room temperature) and with a
microwave (rather than optical) read‑out, our device improves the voltage
sensitivity by roughly three orders of magnitude compared with the
rf-to-optical transducer of Bagci \textit{et al.}~\cite{Bagci2014},
while delivering an output that is natively compatible with quantum
microwave circuitry.

Crucially, the heterodyne mixing of the RF signal with a GHz local oscillator inside the cryogenic cavity bypasses the large Johnson noise of conventional MHz-band amplifiers. The remaining imprecision is set by the noise temperature of the microwave HEMT, operating well below the room-temperature limit. This inherent noise advantage, combined with electrostatic pre-amplification, enables high-sensitivity detection inaccessible to direct rf measurement.

Straightforward improvements would push performance further: shortening the vacuum gap to sub-micrometer dimensions, as demonstrated in \cite{Seis2022}, and adopting ultracoherent Si$_3$N$_4$ drums with $Q>10^{8}$ \cite{Tsaturyan2017,Bereyhi2022,Patange2025} would enhance the electrostatic lever arm by an order of magnitude while suppressing thermal force noise by 20–30~dB. Phononic-crystal defect regions can be engineered with lateral dimensions larger than $200~$\textmu m range \cite{Planz2023,Enzian2023}, and can be functionalized with a metallized (Al) pad that is flip-chip coupled to superconducting capacitor electrodes \cite{Seis2022, Patange2025}. Our model predicts these advances would yield 44~dB total gain and voltage sensitivity below 200~fV/$\sqrt{\mathrm{Hz}}$—rivaling state-of-the-art cryogenic electrometers.

\section*{Acknowledgments}
We thank Antoine Heidmann, Pierre-François Cohadon, and Tristan Briant for insightful discussions. This work was supported by the Agence Nationale de la Recherche under projects \textsc{MecaFlux} (ANR-21-CE47-0011), and \textsc{Ferbo} (ANR-23-CE47-0004).  This work has been supported by Region Île-de-France in the framework of DIM \textsc{Sirteq} (project CryoParis) and DIM QuanTiP (project COCONUT), and by Sorbonne Université through the \textsc{HyQuTech} “Emergence” program. K.~G. acknowledges support from the Quantum Information Center Sorbonne (QICS doctoral fellowship), H.\,P. is funded by the CNRS–University of Arizona joint Ph.D. program.

\bibliographystyle{apsrev4-2}
\nocite{*}
\bibliography{samplebib}

\newpage

\begin{center}
  {\LARGE\bfseries Supplementary Information}\\[0.5em]
  {\large “Cryogenic RF-to-Microwave Transducer based on a DC-Biased 
Electromechanical System”}\\[2em]
\end{center}
\section{Cryostat Wiring}

\begin{figure}[h]
\centering
\includegraphics[width=0.43\linewidth]{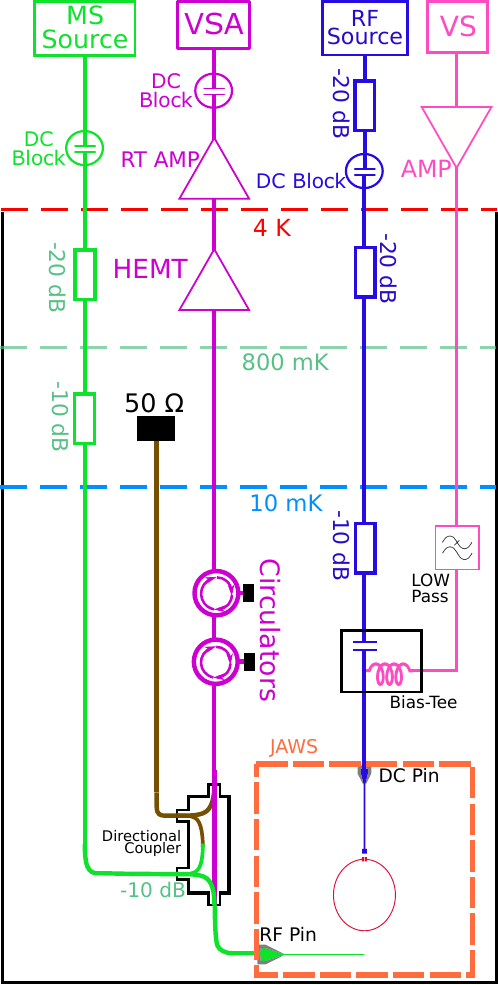}
    \caption{The microwave-shielding sample holder is shown in the 10~mK stage (orange dashed line).  The read-out line input (green line) is connected to a microwave source at room temperature. The signal is attenuated by 20~dB at 4~K and 10~dB at 800~mK. A directional coupler (adding 10~dB attenuation, with power dissipated in a 50~$\Omega$ load at 800~mK),  is then inserted before entering the microwave-shielding box. The reflected signal (purple line) is directed to a double circulator (one port of the circulators is blocked by 50~$\Omega$ loads), and then amplified at 4~K with a HEMT (37 dB gain) and then to a room-temperature amplifier (40~dB gain), before entering a VSA. The rf input signal (blue line) is attenuated by 20~dB at 4~K, and then 10~dB at 10~mK, before entering the dc port of a bias-tee. The voltage source (VS) provides the bias voltage via a high voltage amplifier, which is filtered by a QDevil Qfilter double stage low pass filter (with internal resistance of 1700~$\Omega$) before being connected to the dc port of the bias tee.}
    \label{fig:cryostat}
\end{figure}

The electromechanical and transduction experiments performed in this article have been carried out in a suspended dilution cryostat LD250 from Bluefors at 10~mK temperature. Fig.~\ref{fig:cryostat} shows the cryostat wiring. 

\end{document}